# Mechanical Properties of Schwarzites - A Fully Atomistic Reactive Molecular Dynamics Investigation

Cristiano F. Woellner[1,2], Tiago Botari[1], Eric Perim[1], and Douglas S. Galvão[1,2]

[1]Applied Physics Department, State University of Campinas, 13083-859 Campinas-SP, Brazil

[2]Center for Computational Engineering & Sciences, State University of Campinas, Campinas-SP, Brazil

*Schwarzites are crystalline, 3D porous structures with stable negative curvature formed of sp2-hybridized carbon atoms. These structures present topologies with tunable porous size and shape and unusual mechanical properties. In this work, we have investigated the mechanical behavior under compressive strains and energy absorption of four different Schwarzites, through reactive molecular dynamics simulations, using the ReaxFF force field as available in the LAMMPS code. We considered two Schwarzites families, the so-called Gyroid and Primitive and two structures from each family. Our results also show they exhibit remarkable resilience under mechanical compression. They can be reduced to half of their original size before structural failure (fracture) occurs.*

## INTRODUCTION

Schwarzites are ordered 3D porous graphene-like nanostructures with stable negatively curved sp2-hybridized carbon atoms. These structures were proposed by Mackay and Terrones in 1991. They used the concept of negative curvature in the context of periodic graphitic structures, with the same shapes as triply periodic minimal surfaces (TPMS) [1]. They demonstrated that these structures were energetically stable and therefore possible to synthesize [2,3]. Recently, new porous carbon networks were synthesized [4] and there is a renewed interest in these kind of structures due to their interesting properties and many potential applications, such as energy absorbing materials [5,6]. In this work, we investigated the mechanical behavior under compressive strains of four different Schwarzites through reactive molecular dynamics (MD) simulations. We focus on two Schwarzites families (Gyroid and Primitive) and two structures in each family (see Figures 1-4). These chosen structures differ mainly through their local flatness and they are representative of the large Schwarzite families [1-3].

## THEORY

All MD simulations were carried out using the ReaxFF force field [7], as implemented in the open source code LAMMPS [8,9]. ReaxFF is used in simulations involving classical molecular dynamics but with the advantage of, accurately, describe chemical processes such as formation and breaking of bonds. Its parameterization is obtained using Density Functional Theory (DFT) calculations and its accuracy, compared with experimental data, is around 2.9 kcal/mol for unconjugated and conjugated systems, respectively [7]. It makes ReaxFF potentially applicable to large systems, unlike fully quantum methods. All structures were generated using VMD/TopoTools [10,11].

All structures are symmetric along X, Y and Z directions, so for convenience, we will use Z (**c**-axis) as the strain direction throughout the paper. The compressive response of the structures were calculated decreasing their **c** supercell (simulation box) axis values and allowing their **a** and **b** axes values to freely change consequently decreaing the crystal total volume. Initially, the structures were thermalized at 10 K with a NPT ensemble during 100 ps. This is sufficient time to to the systems reach an equilibrium state. After that, the equilibrated structures were kept at constant temperature (10 K) and allow they can freely expand/contract under compression, we set a zero pressure perpendicular to **c** direction. Periodic boundary conditions were imposed along all the three crytallographic directions.

The compression process is then simulated through a gradual decrease of the **c** lattice parameter. A time step of 0.1 fs was used together with a constant strain rate of $10^{-5}$ per fs. The structural compression is maintained until complete structural failure (fracture). For the structures considered typical simulation time to fracture was in the order of $10^6$ fs.

**DISCUSSION**

In Figures 1 to 4 we present MD snapshots of the different Schwarzites in three different stages of their compressive strain process. The colour in the atoms and bonds represents the local stress (von Mises stress) and goes from darker (low stress) to light (high stress). The first snapshot in each Figure is the equilibrated structure before compression (loading). The following snapshots present the structures at 85, 75 and 50%, respectively, of their initial **c**-values. For these values the structures are already no longer in the linear regime (elastic regime). The observed structural densification takes place through a stacking of 2D graphene-like structures, for all cases. But unlike other 2D structures (i.e., graphene and transition metal dichalcogenides monolayers) in this case these '2D structures' are still covalently bonded. As we can see from these Figure, even for 50% compression, the structure still preserve their structural integrity.

In Figure 5 we present the compressive stress-strain curves (up to failure strain (εF)) of the four Schwarzites considered here. As we can see from this Figure, all structures undergo an elastic/plastic buckling transition. For foams and cellular materials it is common to separate the stress-strain curves into three different regimes: (i) linear (elastic) regime, in this part the structure is well characterized by two constant parameters: Young modulus ($Y_m$) and Poisson ratio (ν); (ii) collapse plateau, in this regime the structures undergo an elastic/plastic buckling and $Y_m$ and ν are no longer constant parameters, but depends on the strain; (iii) densification regime, which is characterized by a rapid increasing in stress values up to the total structural failure (fracture).

From Figures 1-5, comparing the structures in the same family (P688 and P8bal) and (G688 and G8bal) we can identify a similar trend: increasing the ratio of hexagons to octagons the structures become more fracture resistant. It means that they can stand more loading before total structural failure. For both families, the gain is over 20%, see details in Table 1. In Table 1 we summarize some of the mechanical properties of the analysed Schwarzites. We can see from this Table that for all structures the maximum strain supported is over 50%, meaning the structures can be reduced to half of their initial **c**-value before collapsing.

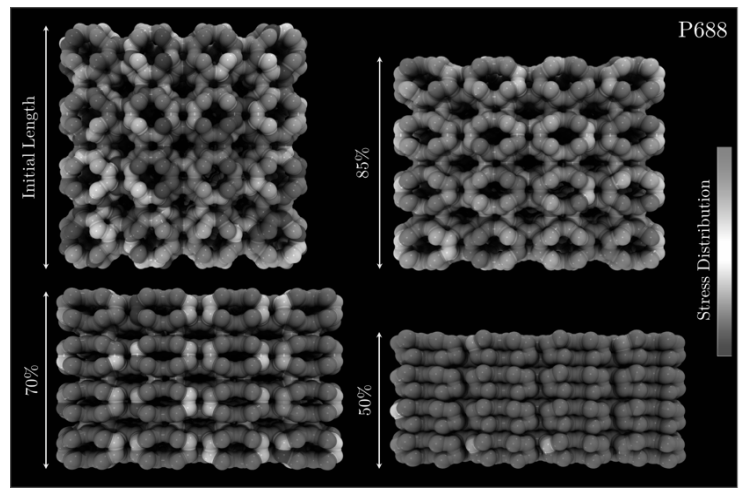

**Figure 1** Representative snapshots from MD simulations of P688 Schwarzite structure under compression in four different stages. The different colours represents the local .

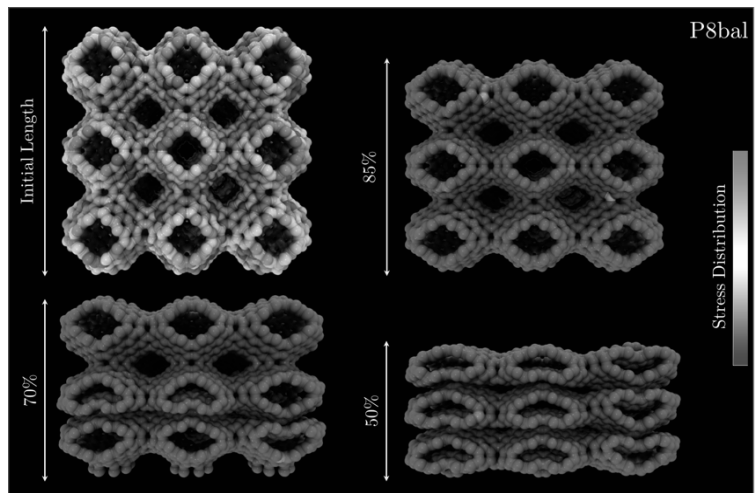

**Figure 2** Representative snapshots from MD simulations of P8bal Schwarzite structure under compression in four different stages. The different colours represents the local stress from low (blue) to high (red) stress value.

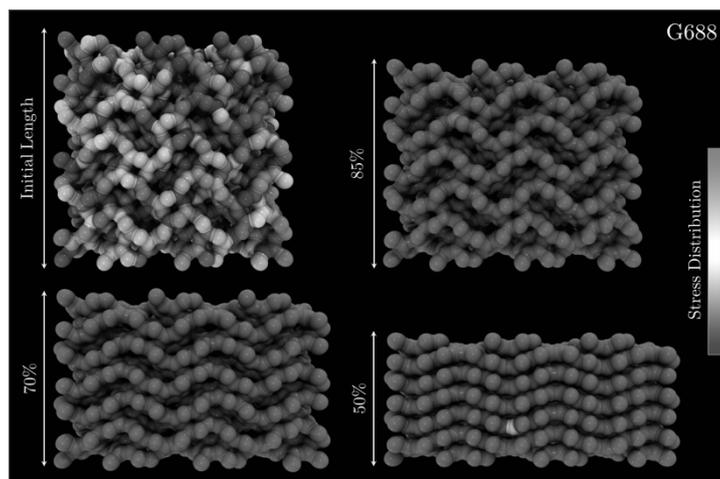

**Figure 3** Representative snapshots from MD simulations of G688 Schwarzite structure under compression in four different stages. The different colours represents the local stress from low (blue) to high (red) stress value. Notice the well-organized stacking

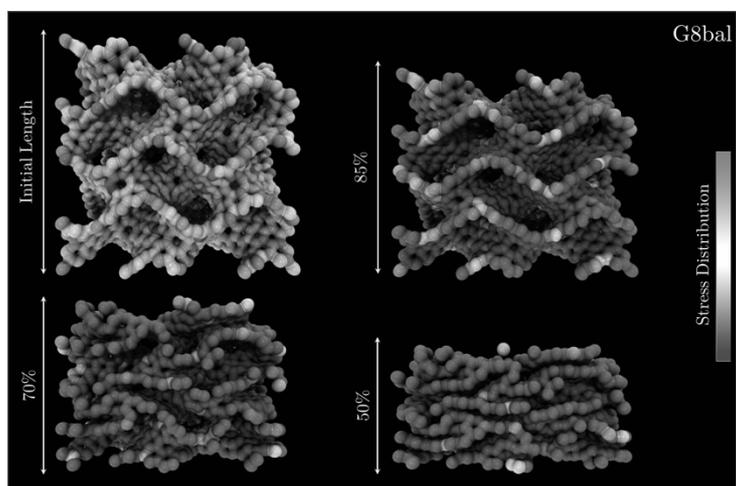

**Figure 4** Representative snapshots from MD simulations of G8bal Schwarzite structure under compression in four different stages. The different colours represents the local stress from low (blue) to high (red) stress value.

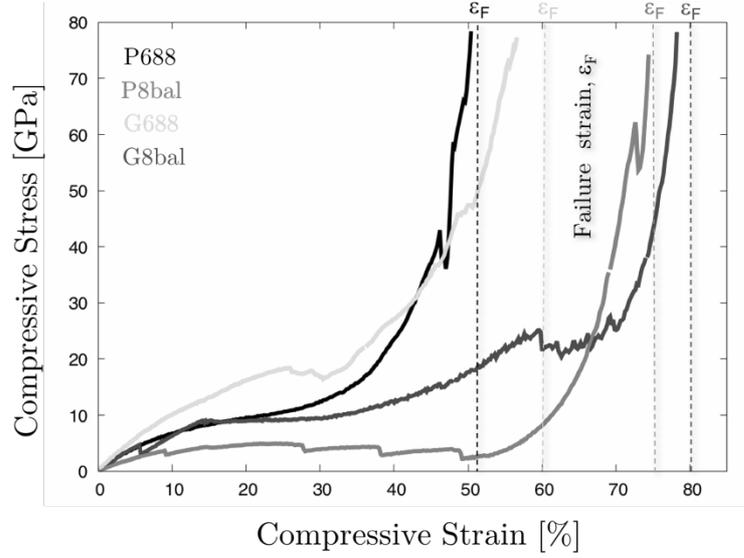

**Figure 5** Compressive stress-strain curves of the four Schwarzites structures under uniaxial compression along the [001] direction up to the failure strain, $\varepsilon_F$.

Another important difference of these structures (P8bal and G8bal) in comparison to (P688 and G688) is the lower stress stored in the structures during the compressive strain. This feature is desirable for applications in energy absorbing materials.

The distinc mechanical behavior of P8bal and G8bal can be explained [2] as follow: increasing the ratio of hexagons in these structures make them locally flatter. Therefore, the in-plane accumulated stress is reduced and lead enhanced their loading resistance.

Table 1. Table of the calculated mechanical properties for all Schwarzites under compression. The columns represent respectively: The structures name, the size of the unit cells, the mass density, the Young modulus, compressive strength (at the compressive failure) and the failure strain.

| Structures | Atoms per unit cell | Mass density (g/cm$^3$) | Young Modulus (GPa) | Compressive Strength (GPa) | Fracture Strain |
|---|---|---|---|---|---|
| **P688**  | 48  | 2.02 | 87.89  | 151.37 | 53.00 |
| **P8bal** | 192 | 1.15 | 51.83  | 116.55 | 76.20 |
| **G688**  | 96  | 2.16 | 121.07 | 340.19 | 59.60 |
| **G8bal** | 384 | 1.22 | 91.42  | 150.40 | 80.20 |

## CONCLUSIONS

We focus on two Schwarzites families and two structures in each family. These chosen structures differ mainly through their local flatness. Our results show that under compression these structures can reduces half of its size before to failure. The feature is considerably enhanced with the increasing of local flatness (more graphenic). Our results also show a with remarkable resilience under compression. These features make these structures good candidates in energy absorbing applications.

## ACKNOWLEDGMENTS

The authors acknowledge the São Paulo Research Foundation (FAPESP) Grant No. 2014/24547-1 for financial support. Computational and financial support from the Center for Computational Engineering and Sciences at Unicamp through the FAPESP/CEPID Grant No. 2013/08293-7 is also acknowledged.

## REFERENCES

1. L. Mackay and H. Terrones, Nature 1991, 352, 762.
2. R. H. Terrones and M. Terrones. N. J. Phys. 2009, 5, 126.
3. D. C. Miller, M. Terrones, and H. Terrones, Carbon, v96, 1191 (2016).
4. K. Kim et al., Nature 2016, 535, 131.
5. L. Yi et al., Carbon 2017, 118, 348.
6. Z. Qin et al., Sci. Adv. 2017, 3, e1601536.
7. S. G. Srinivasan et al., J. Phys. Chem. A 2015, 119, 571.
8. S. Plimpton, J. Comput. Phys. 1995, 117,1.
9. http://lammps.sandia.gov.
10. W. Humphrey et al., J. Molec. Graphics 1996, 14, 33.
11. Axel Kohlmeyer (2017), TopoTools DOI:10.5281/zenodo.545655
12. C. F. Woellner, T. Botari, E. M. Perim, and D. S. Galvao, submitted.